# Mixed convective in an axisymmetric magneto flow owing to $MoS_2$-GO hybrid nanoliquids in $H_2O$ based liquid through an upright cylinder with shape factor

Umair Khan[1], A. Zaib[2], Hamed Daei Kasmaei[3,4*], Manuel Malaver[4,5]

[1]Department of Mathematics and Social Sciences, Sukkur IBA University, Sukkur 65200, Pakistan; Email: umairkhan@iba-suk.edu.pk

[2]Department of Mathematical Sciences, Federal Urdu University of Arts, Science & Technology, Gulshan-e-Iqbal Karachi-75300, Pakistan; Email: aurangzaib@fuuast.edu.pk

[3*]Department of Applied Mathematics, Islamic Azad University-Central Tehran Branch, Tehran, Iran. Email: hamedelectroj@gmail.com (Corresponding Author)
[4]Bijective Physics Institute, Idrija, Slovenia

[4]Bijective Physics Institute, Idrija, Slovenia
[5]Department of Basic Sciences, Maritime University of the Caribbean Catia la Mar, Venezuela. Email: mmf.umc@gmail.com

## Abstract

In the presented paper, a comprehensive study will be done on shape factor analysis of MoS2-GO in H2O-C2H6O2 based hybrid nanoliquids associated with effect and influence of transverse magnetic field and thermal radiation. The effect of variation in different parameters and nanoliquids shapes under temperature and velocity distribution is explored and also non-linear thermal radiation will be analyzed. Algorithms are introduced in proportion to mathematical modeling based on their numerical results and comparative curves for further explanation. In addition, it will be done research for influence and effect of new significant parameters emerged to the model to do sensitivity analysis and also their output results are demonstrated, examined and compared together by presenting graphs and tables. Based on detailed discussions, authentication of attained results designates the high accuracy of applied methods deployed to solve presented model in the paper. Our results satisfy that our used approach is accurate, highly reliable and also effective. All mentioned steps will be described throughout the literature.

**1. Introduction**

In recent years, applications of Nanofluids and their importance has been influenced by many scientists and researchers in the domain of Nanotechnology [1-24, 29-47, 49-53, 59-65]. Nowadays, many technologies and applications have been developed in Nano research centers, universities, organizations, companies, startups and so on. On the other hand, extensive growth of Nanoparticles and their different versions defined by different generations have shown their burgeoning demands in many parts of commercial and industrial development. They have been used to present new solutions and ease processes in Aerospace Engineering [66], ecology [67], biotechnology [68], microbiology [69], Medicine [70,71], Cancer Treatment [72], Genetics [73], Energy [74], electronic devices [75] and so on. Nano fluids (NFs) were made and introduced to the world of science and technology for the first time by Choi [48] in Argonne National Laboratory and their main goal was to explain fluids. Nanoscale elements include metallic substances such as aluminium (Al), copper (Cu), silver (Ag) and Titanium (Ti) associated with oxides and without their oxides. When they have been considered and formed without their oxides, then they can be considered as a kind of very technical importance as well as solid particles that is merged with other ingredients and various molecular compounds. When nanoscale elements are added to a base fluid or molecular fluid, thermal performance and conductivity of Nano-liquids are evaluated to enrich them double in longer surface area.

Nanoparticles applied in Nanofluids have been formed of metals (AL, Ag) and (Cu) oxides (Al2O3, TiO2), nitrides (AlN, SiN), carbides (SiC), or nonmetals (carbon nanotubes, graphite). They consist of solid Nano particles such that their size is less than 100 nm with solid volume fractions typically less than 4% and their diameter is considered between 1 to 100 nm and its molecular fluid structure is similar to ethylene glycol (C2H6O2), water (H2O), engine oil, lubricants and other similar components. Efficiency and effectiveness of Nano liquids and their applications, reliability and their contribution based on magneto hydrodynamic (MHD) flows has been proved in nanotechnology in which they enrich and make progress in process of thermal and convective heat transfer under a boundary layer flow. Some of applications of Nano liquids in different parts of Technology and Industry can be found in [25-28].

Aluminum Oxides that are known as chemical combination (Al2O3) or Alumina can be counted for a part of Nanoscale elements that is applied in order to make rigorous thermal quality performance of Molecular liquids. They have been classified into two types that are known as αAl2O3 or γAl2O3 based on their magnitudes. Phenomenon of bunch scattering on the stretching cylinder through Al2O3 and Cu−water-Nanofluids has been studied by Alshomri and Gui [54]. The entropy generation and 2D stream property in γAl2O3−H2O and γAl2O3−C2H6O2 Nanoliquids by Prandtl number model has been investigated in [55].3D flow state in γAl2O3−H2O and γAl2O3−C2H6O2 Nanoliquids amid parallel rotating surfaces has been interrogated by Khan et al [56]. Also, this type of Nanoliquids from aspect of entropy analysis of model has been worked by Hayat et al [57].Chemical reaction and Thermal radiation of MHD streaming of Nanoliquids and heat transport investigation on Water based Nanofluids coated with Ag, TiO2, Cu and Al2O3 on a stretchable sheet has been surveyed by Jain et al [58]. Now, we will explain full structure of paper briefly.

In this paper, at first, we define all notions and parameters that are used in different equations. Then, we explain about formulation and modelling of Nano liquids. At next step, we use finite difference scheme in Matlab Software (bvp4c Package) to solve the obtained model. Then we implement simulation process based on parameter estimation in different states. Afterwards, results and discussion will be done about obtained results from Simulation and at the end, we conclude about convergence of results based on implemented method as a reliable technique.

## 2. Formulation of the problem

Consider a steady problem of mixed convection axisymmetric stagnation point flow through an upright stretching cylinder as shown in Fig. 1. Shape factor, radiation and magnetic fields are taken into account in this perusal. The cylinder surface is in contact through involving of $MoS_2$-GO hybrid nano-liquids in water-based liquid. The cylinder is expressed by $r = a$ in the polar cylindrical coordinate. The flow is axisymmetric regarding the $z-$axis and also symmetric at $z = 0$ plane. The ambient liquid has an invariable temperature $T_\infty$ and cylinder is preserved at a temperature $T = T_w$. Having $T_w < T_\infty$ signifies the opposing flow, while $T_w > T_\infty$ indicating the assisting flow. The governing equations are expressed as

$$\frac{\partial(ru)}{\partial r} + r\frac{\partial w}{\partial z} = 0 \tag{1}$$

$$u\frac{\partial u}{\partial r} + w\frac{\partial u}{\partial z} = -\frac{1}{\rho_{hbnf}}\frac{\partial p}{\partial r} + v_{hbnf}\left(\frac{\partial^2 u}{\partial r^2} + \frac{1}{r}\frac{\partial u}{\partial r} - \frac{u}{r^2}\right) - \frac{\sigma_{hbnf} B_0^2}{\rho_{hbnf}} u \tag{2}$$

$$u\frac{\partial w}{\partial r} + w\frac{\partial w}{\partial z} = -\frac{1}{\rho_{hbnf}}\frac{\partial p}{\partial z} + v_{hbnf}\left(\frac{\partial^2 w}{\partial r^2} + \frac{1}{r}\frac{\partial w}{\partial r}\right) - \frac{\sigma_{hbnf} B_0^2}{\rho_{hbnf}} w \pm g\frac{(\rho\beta)_{hbnf}}{\rho_{hbnf}}(T - T_\infty) \tag{3}$$

$$u\frac{\partial T}{\partial r}+w\frac{\partial T}{\partial z}=\frac{k_{hbnf}}{(\rho c_p)_{hbnf}}\left(\frac{\partial^2 T}{\partial r^2}+\frac{1}{r}\frac{\partial T}{\partial r}\right)-\frac{1}{(\rho c_p)_{hbnf}}\frac{1}{r}\frac{\partial}{\partial r}(rq_r) \qquad (4)$$

The corresponding boundary restrictions are

$$\begin{aligned} &u=0,\ w=2Az,\ T=T_w=T_\infty+A_1 z\ \text{at}\ r=a,\\ &u=-B\left(r-\frac{a^2}{r}\right),\ w=2Bz,\ T\to T_\infty\ \text{as}\ r\to\infty. \end{aligned} \qquad (5)$$

$z$ and $r$ are the cylindrical polar-coordinates distances measuring in the axial and radial directions, respectively. The magnetic field, the pressure, the acceleration are identified as $B_0$, $p$ and $g$, respectively. Moreover, $\nu_{hbnf}$, $\rho_{hbnf}$, $\beta_{hbnf}$, $\sigma_{hbnf}$, $k_{hbnf}$, $(\rho c_p)_{hbnf}$ are respectively symbolizes kinematic viscosity, density, thermal expansion coefficient, electrical conductivity, thermal conductivity and heat capacity of the hybrid nanoliquids. The final term in Eq. (3) represents the effect of buoyancy force has sign $\pm$, where the positive sign indicates the buoyancy assisting flow while negative sign indicates the buoyancy opposing flow. In Eq. (4), $q_r$ is the radiative heat flux which is expressed through the approximation of the Rosseland

$$q_r=-\frac{16\sigma^* T_\infty^3}{3k^*}\frac{\partial T}{\partial r} \qquad (6)$$

where $k^*$ signifies the mean constant of absorption and $\sigma^*$ indicates the constant of Stefan Boltzmann.

Here, we endeavored to attempt a novel method to alter the mode of heat transfer in fluids which are freshly examined among researchers. We utilized the hybrid nanomaterials together with varied structure of nanomaterial and base fluid. The present way is estimated to be a proficient path to modify the process of transfer rate of heat in fluids. Therefore, the attributes of thermo-physical feature of base fluid and hybrid nanofluids are given in Table 1 with different shape effects *m* that is suggested in Table 2. Whereas, Table 3 shows the values of thermo physical properties of the regular fluid and nanoparticles.

We introduced the following similarity transformations:

$$u = -Aa\frac{f(\xi)}{\sqrt{\xi}},\ w = 2Azf'(\xi),\ \xi = \left(\frac{r}{a}\right)^2, \theta(\xi) = \frac{T-T_\infty}{T_w - T_\infty}. \tag{7}$$

Implementing the above transformations into Eqs. (2) to (5) and via the well-known equation (7) in order to obtain the consistent transmuted ODE's, we get:

$$g_1(\xi f''' + f'') + g_2 \text{Re}_a\left(\alpha^2 + ff'' - (f')^2\right) + g_3 \text{Re}_a M(\alpha - f') + g_4 \lambda \text{Re}_a \theta = 0 \tag{8}$$

$$(\xi\theta'' + \theta')\left[g_5 + \frac{4}{3}R_d\right] + g_6 \text{Re}_a \Pr(f\theta' - f'\theta) = 0 \tag{9}$$

The subjected major boundary restriction is

$$f'(1) - 1 = 0,\ f(1) = 0,\ \theta(1) - 1 = 0,$$
$$f'(\infty) - \alpha \to 0,\ \theta(\infty) \to 0. \tag{10}$$

In which:

$$g_1 = \frac{1}{(1-\phi_1)^{2.5}(1-\phi_2)^{2.5}},$$

$$g_2 = \left((1-\phi_2)\left\{(1-\phi_1)+\phi_1\frac{\rho_{s_1}}{\rho_f}\right\}+\phi_2\frac{\rho_{s_2}}{\rho_f}\right),$$

$$g_3 = \left(\left[\frac{\sigma_{s_2}(1+2\phi_2)+2\sigma_{bf}(1-\phi_2)}{\sigma_{s_2}(1-\phi_2)+\sigma_{bf}(2+\phi_2)}\right]\right)\left(\frac{\sigma_{s_1}(1+2\phi_1)+2\sigma_f(1-\phi_1)}{\sigma_{s_1}(1-\phi_1)+\sigma_f(2+\phi_1)}\right),$$

$$g_4 = \left((1-\phi_2)\left[(1-\phi_1)+\phi_1\frac{(\rho\beta)_{s_1}}{(\rho\beta)_f}\right]+\phi_2\frac{(\rho\beta)_{s_2}}{(\rho\beta)_f}\right),$$

$$g_5 = \left\{\frac{k_{s_2}-k_{nf}(1-m)+\phi_2(1-m)(k_{nf}-k_{s_2})}{k_{s_2}-k_{nf}(1-m)+\phi_2(k_{nf}-k_{s_2})}\right\}\left\{\frac{k_{s_1}-k_f(1-m)+\phi_1(1-m)(k_f-k_{s_1})}{k_{s_1}-k_f(1-m)+\phi_1(k_f-k_{s_1})}\right\},$$

$$g_6 = \left((1-\phi_2)\left[\phi_1\frac{(\rho c_p)_{s_1}}{(\rho c_p)_f}+(1-\phi_1)\right]+\phi_2\frac{(\rho c_p)_{s_2}}{(\rho c_p)_f}\right),$$

(11)

The non-dimensional constraints in Eqs. (8) - (9) and (10) are mathematically expressed as:

$$\lambda = \frac{Gr_z}{\text{Re}_z^2}, M = \frac{\sigma_f B_0^2}{2\rho_f A}, \text{Re}_z = \frac{Az^2}{2v_f}, \Pr = \frac{v_f}{\alpha_f}, Gr_z = \frac{g\beta_f(T_w-T_\infty)z^3}{16v_f^2}, \text{Re}_a = \frac{Aa^2}{2v_f}, \alpha = \frac{B}{A},$$

$$R_d = \frac{4\sigma^* T_\infty^3}{k_f k^*}.$$

while respectively shows their proper names of these aforementioned parameters that is conspicuously exercise in the problem called the mixed convective parameter $(\lambda)$ (where it is defined as $\lambda = \frac{Gr_z}{\text{Re}_z^2}$ called the fraction of $(Gr_z)$ Grashof number and the Reynolds number $(\text{Re}_z)$, magnetic parameter $(M)$, Prandtl number $(\Pr)$, free stream Reynolds number $(\text{Re}_a)$, ratio parameter and $(R_d)$ radiation parameter.

The local Nusselt and the skin friction factor are the physical significant quantities concerning on the flow with heat transport. These quantities in the ODE's form are

$$Nu_z = \frac{-k_{hbnf} z}{k_f (T_w - T_\infty)} \left(\frac{\partial T}{\partial r}\right)_{r=a} \tag{12}$$

$$C_{fz} = \frac{\mu_{hbnf}}{\frac{1}{2}\rho w^2}\left(\frac{\partial w}{\partial r}\right)_{r=a}, w = 2Az \tag{13}$$

Applying (7) into (12) and (13), we get:

$$\frac{1}{2}\left(\frac{\operatorname{Re}_z}{\operatorname{Re}_a}\right)^{-\frac{1}{2}} Nu_z = \frac{-k_{hbnf}}{k_f} \theta'(1) \tag{14}$$

$$\sqrt{\operatorname{Re}_z \operatorname{Re}_a}\, C_{fz} = \frac{f''(1)}{(1-\phi_1)^{2.5}(1-\phi_2)^{2.5}} \tag{15}$$

## 3. Results and discussions

In this portion, the nonlinear ODE's (8) and (9) with restricted condition (10) has been worked out numerically through a solver bvp4c. The outcomes of diverse constraints in the presence of shape factor of nanoparticle on liquid velocity, temperature profile along with friction factor and rate of heat transfer for nanoliquids as well as hybrid nanoliquid phases have been examined in the form of graphs (Figs. 2-13) as well as tabulated in the form of Tables (Tables 5-6). Also, assisting and opposing flows were discussed. The range of constraints in this research are considered as: $0.01 \leq \lambda \leq 0.2$, $0 \leq M \leq 4$, $0 \leq \alpha \leq 1.5$, $0 \leq R_d \leq 1$, $1 \leq \operatorname{Re}_a \leq 3$, $0 \leq \phi_1 \leq 0.1$ and $0 \leq \phi_2 \leq 0.003$. Table 4 is prepared for validation of the current result $f''(1)$ with published result of Fang et al. [1] and Hamid et al. [2] in the limiting cases and found an excellent agreement.

Figs. 2 and 3 are set to inspect the influence of mixed parameter $\lambda$ on the velocity and temperature profiles. It is transparent from Fig. 2 that the velocity of liquid is more pronounce for greater $\lambda$ in case of assisting flow. Physically, greater amount of $\lambda$ generated great buoyancy force which gives the highest moving energy and as a result, such energy generated the confrontation through the flow. The contrary trend is scrutinized for the velocity in opposing flow case. Fig. 3 reveals that the temperature and the corresponding boundary layer decline due to $\lambda$ in the assisting and opposing flows. Figs. 4 and 5 depict the influence of magnetic parameter on the velocity profile and temperature distribution. Fig. 4 indicates that magnetic parameter resists with the velocity of liquid in the hybrid nanoliquid as well as nanoliquid and as a result, the velocity boundary-layer thickness declines. Physically, the presence of magnetic field generates the Lorentz forces that are in fact the drag force. The flow and Lorentz force act in contrary track to each other, comprising a flow retardation influence. Fig. 5 explains that by amplifying the potency of magnetic parameter, the liquid temperature augments. Physically, the resistive kind of force known as Lorentz force counters with the liquid motion, thus heat is fabricated and consequently temperature and corresponding boundary-layer thickness developed into thicker. In addition, it is clear from these profiles that the liquid flow accelerates more for hybrid nanoparticles as compared to $MoS_2$/water nanoparticle. Figs. 6 and 7 highlight the impact of radiation parameter on the fluid velocity and temperature distribution. It is transparent from these portraits that the velocity and temperature of liquid augment due to magnifying the radiation parameter. Physically, the surface of heat flux increases by the radiation and consequently larger temperature in the boundary-layer region should be approximated. The achieved result is a confirmation of the legitimacy of the relation $R_d$. Also, the radiation is utilized to collapse the molecules of water into hydrogen. The impacts of shape factors on the temperature distribution and velocity are portrayed in Figs. 8 and 9. The velocity and temperature of the hybrid as well as nanofluids augment with shape factor. Physically, sturdy hydrogen bonding of hybrid nanoliquid and nanoliquid cause a sharp augment in the thermal conductivity and thus the velocity and temperature profiles enhance. In addition, the temperature is maximum for the blade shape and minimum in the case of cylinder for hybrid nanoliquid as well as for nanoliquid. Figs. 10 and 11 are demonstrated as the inspiration of $\text{Re}_a$ on the temperature and velocity fields for hybrid nanoliquid and nanoliquid. It is expected from these profiles that the velocity as well as the temperature declines with rising the values of $\text{Re}_a$. The impact of nanoparticle volume fraction $\phi_2$ on the velocity and the temperature profiles are depicted in Figs. 12 and 13 for the assisting and the opposing flows. Fig, 12 explains that the velocity augments with $\phi_2$ for $\lambda > 0$ and declines for $\lambda < 0$. The upsurge in the velocity is owing to the reality that dynamic viscosity of hybrid nanoliquid has inverse relation with volume fraction. Therefore, an augment in $\phi_2$ guides to decline the viscosity of regular liquid and consequently accelerates the liquid flow. Whereas, the contrary impact is seen on the temperature (Fig. 13).

Tables 5 and 6 are prepared to see the influence of volume fraction $\phi_1$ and magnetic parameter $M$ on the friction factor and the rate of heat transfer for hybrid nanoliquid and nano with different shape factors, respectively. It is apparent from Table 1 that the friction factor and the heat transfer rate augment with $\phi_1$ for the hybrid nanoliquid and nanoliquid. Physically, the thermal conductivity is enhanced due to $\phi_1$, which consequently boost up the rate of heat transfer in both nanoliquids. In addition, the friction factor and the rate of heat transfer are greater in the case of blade shape and lower in the cylinder shape. In Table 2, the values of the friction factor and the heat transfer rate decline with augmenting $M$ for hybrid Nanoliquid and Nanoliquid.

## 4. Conclusion

In this perusal, the mixed convective magneto flow with heat transfer containing $MoS_2$-GO/water hybrid nanoliquids near a stagnation point through a vertical stretched cylinder with shape factor and radiation impact have been explored. The similarity technique is employed to alter the PDE's into nonlinear ODE's and these transmuted PDE's are worked out through bvp4c solver. The significant outcomes are summarized as:

- The liquid velocity upsurges with augmenting $\lambda$ for the assisting flow and declines in the opposing flow. While, the temperature distribution shrinks in the assisting and opposing flows.
- Due to the presence of magnetic field, the velocity declines and temperature augments for hybrid nanoliquid and nanoliquid.
- The velocity and the temperature show increasing behavior due to radiation parameter.
- The influence of shape factors has an incremental and assenting effect on the velocity and the temperature profiles as well as on the skin factor and the Nusselt number.
- The impact of Reynolds number on the velocity and temperature behaves in decreasing way for both nanoliquids.
- The velocity increases due to $\phi_1$ in the assisting flow and decelerated in the opposing flow, while the opposite impact is observed on the temperature profile.
- The Nusselt number and the skin factor decline due to $M$.

**Table 1:** Thermo-physical attributes of hybrid nanofluid and regular fluid.

| Properties | Nanofluid | Hybrid nanofluid |
|---|---|---|
| Density | $\rho_{nf} = \{(1-\phi)\rho_f + \phi\rho_s\}$ | $\rho_{hbnf} = \left[(1-\phi_2)\{(1-\phi_1)\rho_f + \phi_1\rho_{s_1}\} + \phi_2\rho_{s_2}\right]$ |
| Viscosity | $\mu_{nf} = \dfrac{\mu_f}{(1-\phi)^{2.5}}$ | $\mu_{hbnf} = \dfrac{\mu_f}{(1-\phi_1)^{2.5}(1-\phi_2)^{2.5}}$ |
| Thermal expansion | $(\rho\beta)_{nf} = \left[(1-\phi)(\rho\beta)_f + \phi(\rho\beta)_s\right]$ | $(\rho\beta)_{hbnf} = (1-\phi_2)\left[(1-\phi_1)(\rho\beta)_f + \phi_1(\rho\beta)_{s_1}\right] + \phi_2(\rho\beta)_{s_2}$ |
| Electrical conductivity | $\sigma_{nf} = \sigma_f\left[1 + \dfrac{3(\sigma-1)\phi}{(\sigma+2)-(\sigma-1)\phi}\right]$ | $\sigma_{hbnf} = \sigma_{bf}\left[\dfrac{\sigma_{s_2}(1+2\phi_2) + 2\sigma_{bf}(1-\phi_2)}{\sigma_{s_2}(1-\phi_2) + \sigma_{bf}(2+\phi_2)}\right]$ with $\sigma_{bf} = \sigma_f\left[\dfrac{\sigma_{s_1}(1+2\phi_1) + 2\sigma_f(1-\phi_1)}{\sigma_{s_1}(1-\phi_1) + \sigma_f(2+\phi_1)}\right]$ |
| Thermal conductivity | $\dfrac{k_{nf}}{k_f} = \dfrac{k_s + (m-1)k_f - (m-1)\phi(k_f - k_s)}{k_s + (m-1)k_f + \phi(k_f - k_s)}$ | $\dfrac{k_{hbnf}}{k_{bf}} = \dfrac{k_{s_2} + (m-1)k_{bf} - (m-1)\phi_2(k_{bf} - k_{s_2})}{(\hat{k}_{s_2} + 2\hat{k}_{nf}) + \phi_2(\hat{k}_{nf} - \hat{k}_{s_2})}$ with $k_{bf} = \dfrac{k_{s_1} + (m-1)k_f - (m-1)\phi_1(k_f - k_{s_1})}{k_{s_1} + (m-1)k_f - \phi_1(k_f - k_{s_1})} \times k_f$ |
| Heat capacity | $(\rho c_p)_{nf} = \left[(1-\phi)(\rho c_p)_f + \phi(\rho c_p)_s\right]$ | $(\rho c_p)_{hbnf} = (1-\phi_2)\left[(1-\phi_1)(\rho c_p)_f + \phi_1(\rho c_p)_{s_1}\right] + \phi_2(\rho c_p)_{s_2}$ |

**Table 2:** Shapes of the nanoparticle with their values.

| Shapes of the factors | | Shape factor |
|---|---|---|
| Bricks | 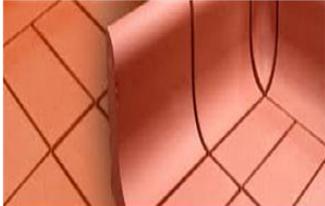 | 3.7 |
| Cylinders | 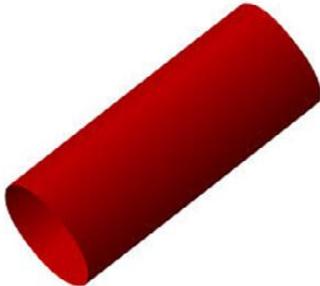 | 4.9 |
| Platelets | 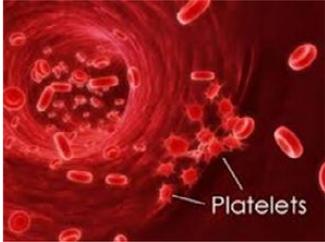 | 5.7 |
| Blades | 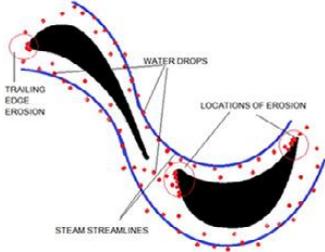 | 8.6 |

**Table 3:** Thermo physical properties of the base fluid and hybrid nanoparticles [Ghadikolaei and Gholinia [3]].

| Characteristic properties | $H_2O$ | $MoS_2$ | GO |
|---|---|---|---|
| $\rho$ | 997.1 | 5060 | 1800 |
| $c_p$ | 4179 | 397.21 | 717 |
| $k$ | 0.613 | 904.4 | 5000 |
| $\sigma$ | 0.005 | $2.09 \times 10^4$ | $6.30 \times 10^7$ |
| $\beta$ | 21 | $2.8424 \times 10^{-5}$ | $2.84 \times 10^{-4}$ |
| Pr | 6.2 | - | - |

**Table 4:** Comparison of current result $[f''(1)]$ with existing result in steady flow case when $M = \alpha = \text{Re}_a = R_d = \lambda = \phi_1 = \phi_2 = 0$.

| Current result | Fang et al. [1] | Hamid et al. [2] |
|---|---|---|
| -1.1778 | -1.17775 | -1.17776 |

**Table 5:** Impact of nanoparticle fraction $\phi_1$ on $\sqrt{\text{Re}_z \text{Re}_a} C_{fz}$ and $(1/2)(\text{Re}_z/\text{Re}_a)^{\frac{-1}{2}} Nu_z$ for $MoS_2$\water and $MoS_2$–GO/water when $M = 0.5, \alpha = 1, \text{Re}_a = 1, R_d = 0.5, \lambda = 0.01, \phi_2 = 0.001$.

| | $\phi_1$ | 3.7 | | 4.9 | | 5.7 | | 8.6 | |
|---|---|---|---|---|---|---|---|---|---|
| | | Nano | Hybrid | Nano | Hybrid | Nano | Hybrid | Nano | Hybrid |
| $\sqrt{\text{Re}_z \text{Re}_a} C_{fz}$ | 0.01 | 0.1154 | 1.2700 | 0.1157 | 1.2729 | 0.1159 | 1.2747 | 0.1165 | 1.2815 |
| | 0.02 | 0.2295 | 1.3855 | 0.2305 | 1.3916 | 0.2312 | 1.3955 | 0.2337 | 1.4097 |
| | 0.03 | 0.3445 | 1.5026 | 0.3468 | 1.5122 | 0.3484 | 1.5185 | 0.3537 | 1.5406 |
| $\frac{1}{2}\left(\frac{\text{Re}_z}{\text{Re}_a}\right)^{\frac{-1}{2}} Nu_z$ | 0.01 | 2.7240 | 2.8377 | 2.7469 | 2.8622 | 2.7620 | 2.8784 | 2.8164 | 2.9368 |
| | 0.02 | 2.8006 | 2.9138 | 2.8461 | 2.9623 | 2.8762 | 2.9944 | 2.9835 | 3.1089 |
| | 0.03 | 2.8773 | 2.9901 | 2.9454 | 3.0624 | 2.9902 | 3.1100 | 3.1491 | 3.2791 |

**Table 6:** Impact of magnetic parameter $M$ on $\sqrt{Re_z Re_a} C_{fz}$ and $(1/2)(Re_z/Re_a)^{\frac{-1}{2}} Nu_z$ for MoS$_2$\water and MoS$_2$–GO/water when $\phi = 0.01, \alpha = 1, Re_a = 1, R_d = 0.5, \lambda = 0.01, \phi_2 = 0.001$.

|  | $M$ | 3.7 | | 4.9 | | 5.7 | | 8.6 | |
|---|---|---|---|---|---|---|---|---|---|
|  |  | Nano | hybrid | Nano | hybrid | Nano | hybrid | Nano | hybrid |
| $\sqrt{Re_z Re_a} C_{fz}$ | 0.0 | 0.1189 | 1.3024 | 0.1192 | 1.3054 | 0.1194 | 1.3074 | 0.1200 | 1.3145 |
|  | 0.5 | 0.1154 | 1.2700 | 0.1157 | 1.2729 | 0.1159 | 1.2747 | 0.1165 | 1.2815 |
|  | 1.0 | 0.1123 | 1.2409 | 0.1126 | 1.2436 | 0.1128 | 1.2454 | 0.1134 | 1.2519 |
| $\frac{1}{2}\left(\frac{Re_z}{Re_a}\right)^{\frac{-1}{2}} Nu_z$ | 0.0 | 2.7246 | 2.8427 | 2.7475 | 2.8673 | 2.7626 | 2.8836 | 2.7418 | 2.8627 |
|  | 0.5 | 2.7240 | 2.8377 | 2.7469 | 2.8622 | 2.7620 | 2.8784 | 2.7412 | 2.8576 |
|  | 1.0 | 2.7235 | 2.8332 | 2.7463 | 2.8576 | 2.7615 | 2.8739 | 2.7406 | 2.8530 |

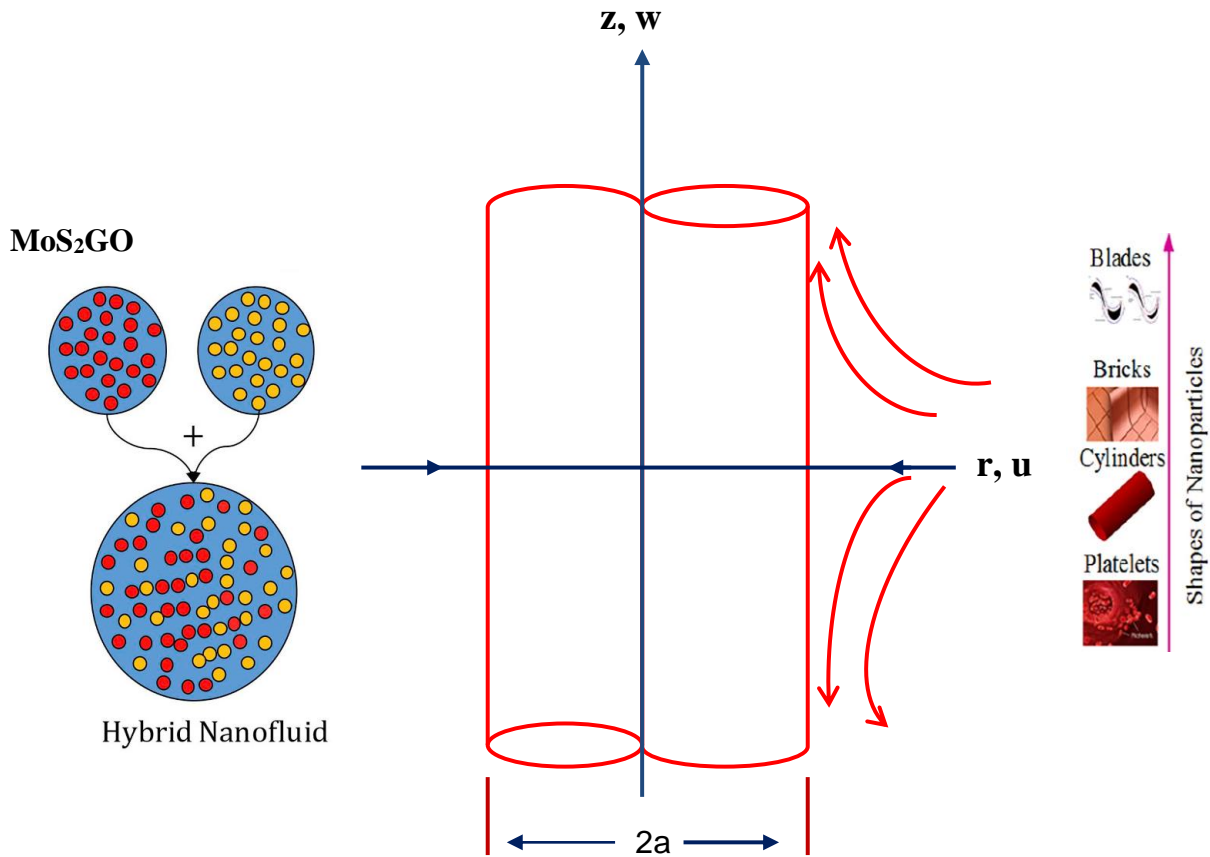

**Fig. 1**: Physical diagram of the problem [3].

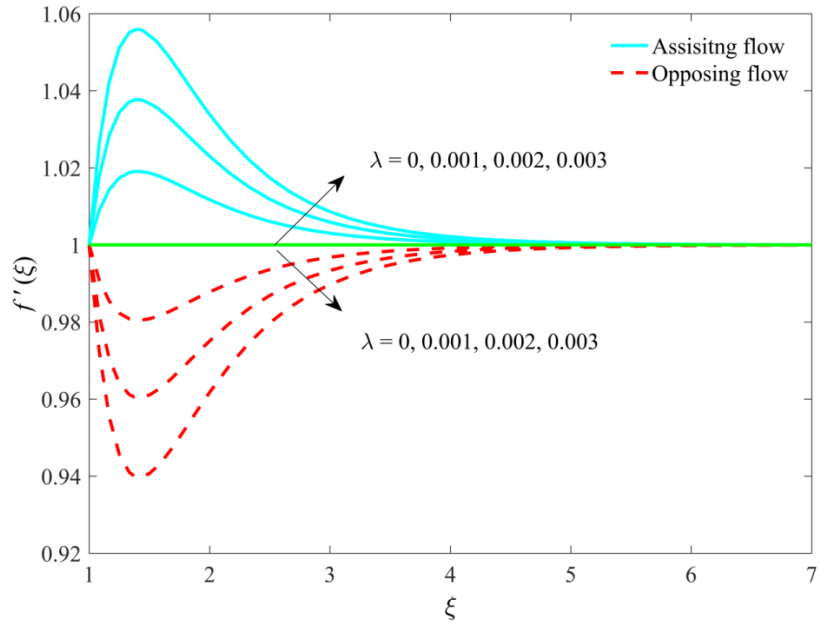

**Fig. 2:** Impact of $\lambda$ on $f'(\xi)$.

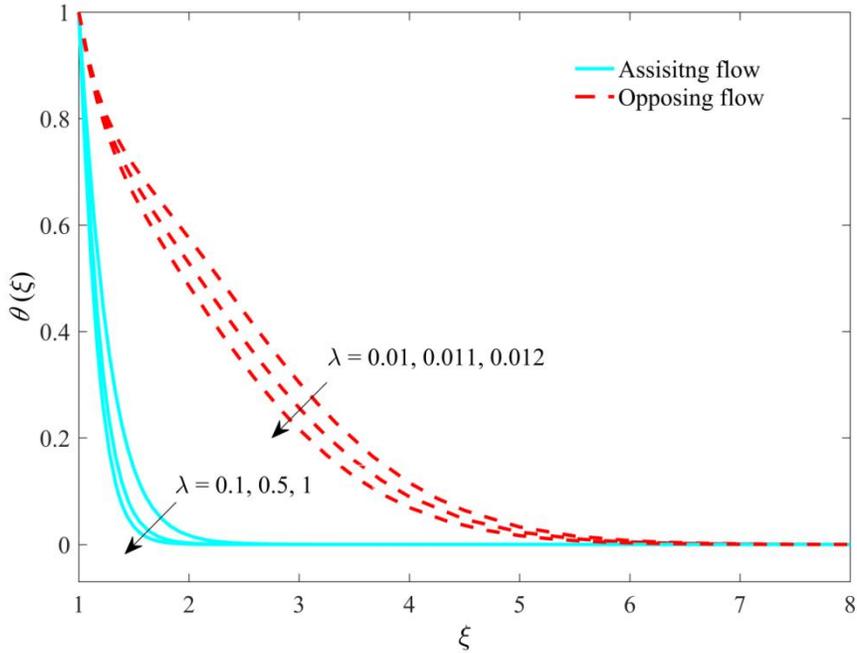

**Fig. 3:** Impact of $\lambda$ on $\theta(\xi)$.

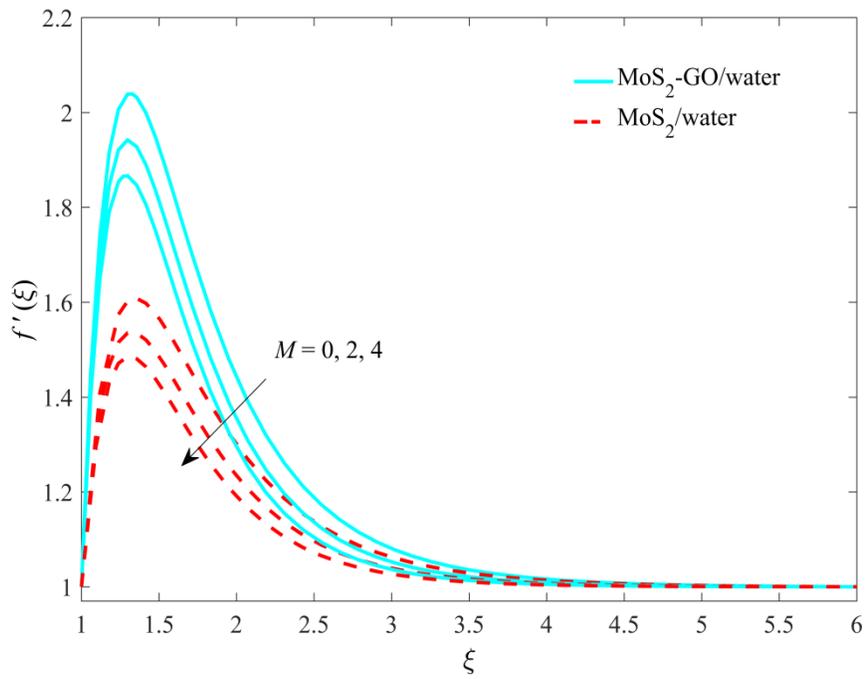

**Fig. 4:** Impact of $M$ on $f'(\xi)$.

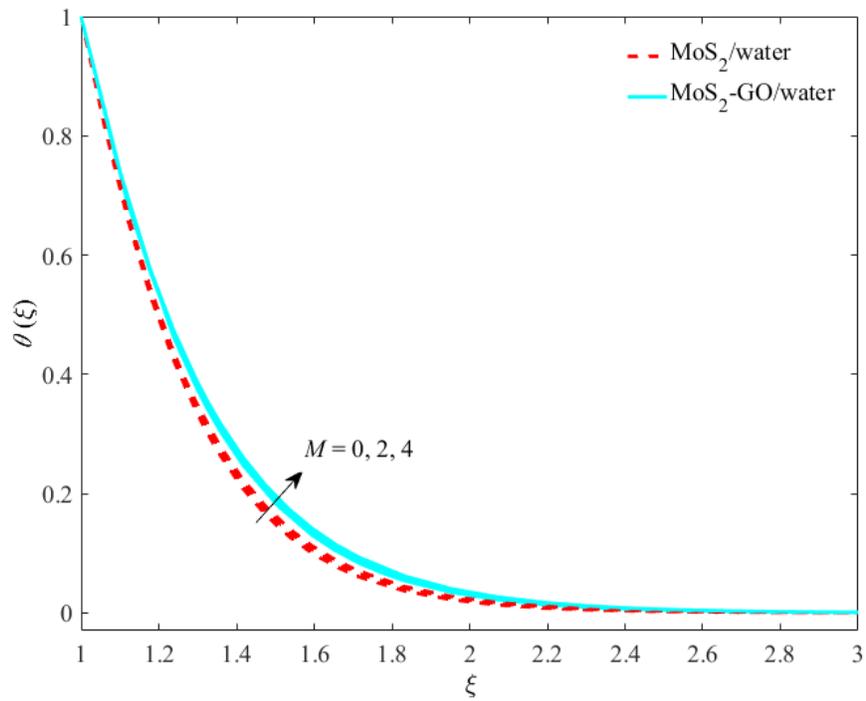

**Fig. 5:** Impact of $M$ on $\theta(\xi)$.

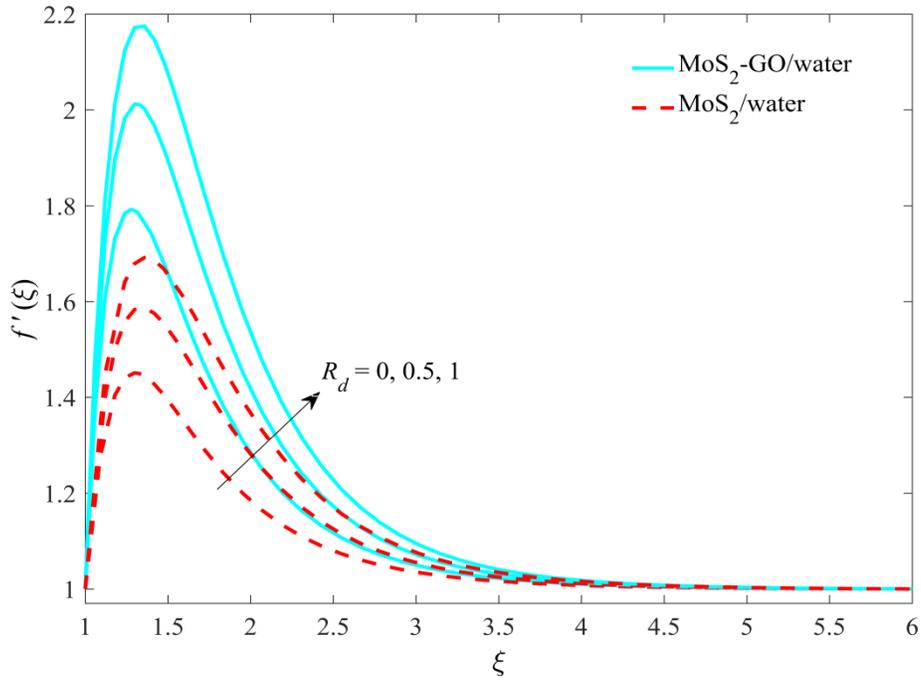

**Fig. 6:** Impact of $R_d$ on $f'(\xi)$.

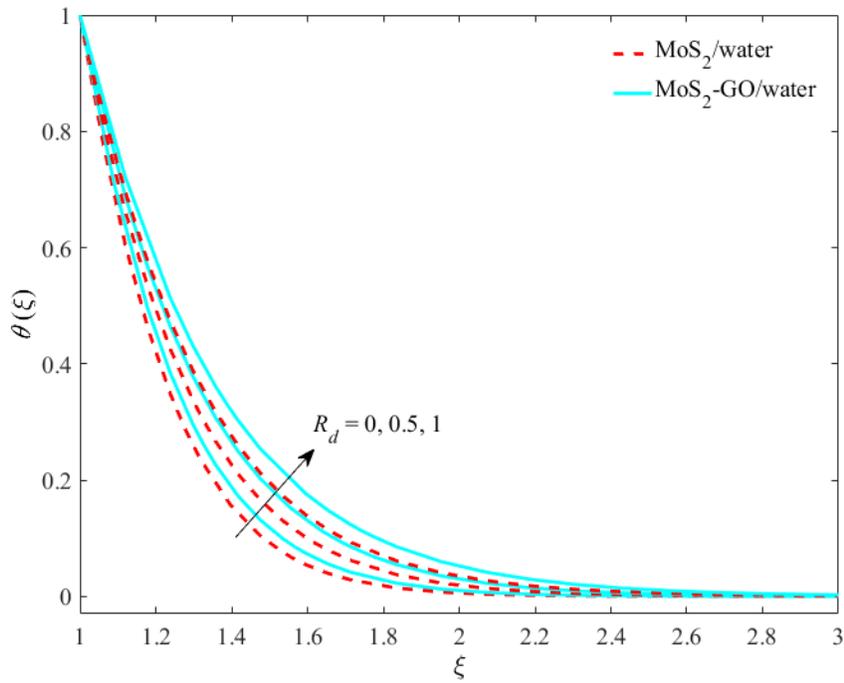

**Fig. 7:** Impact of $R_d$ on $\theta(\xi)$.

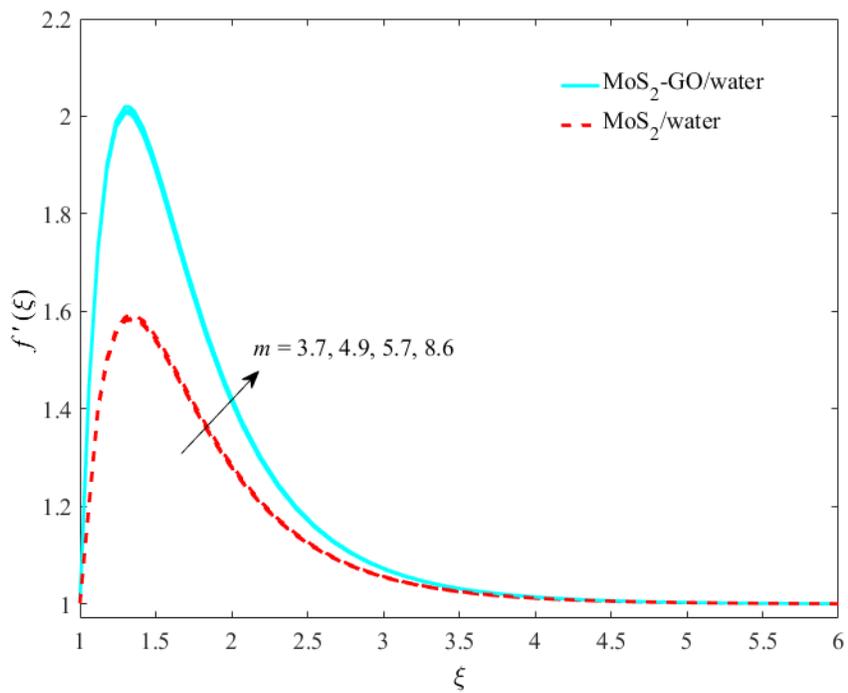

**Fig. 8:** Impact of $m$ on $f'(\xi)$.

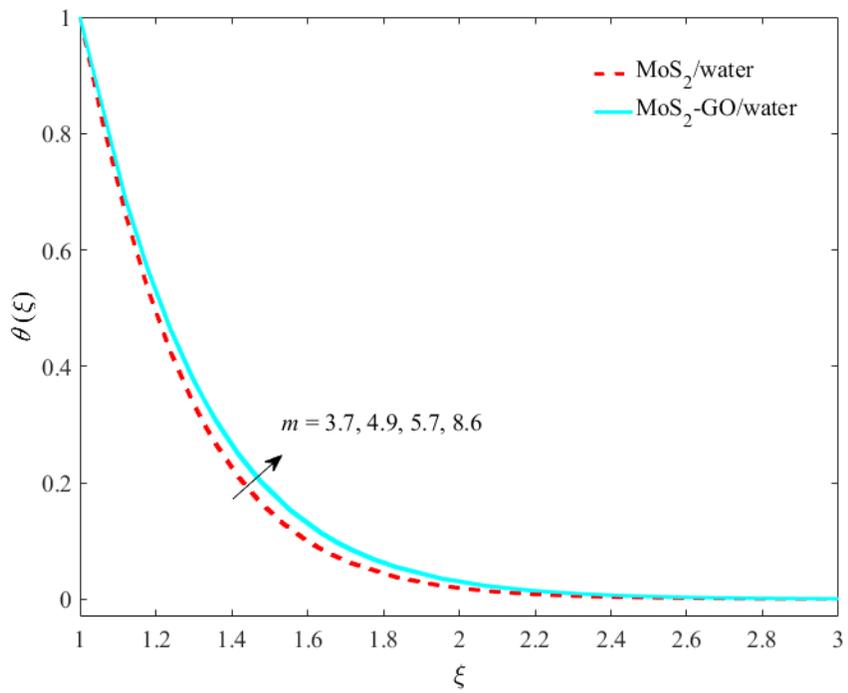

**Fig. 9:** Impact of $m$ on $\theta(\xi)$.

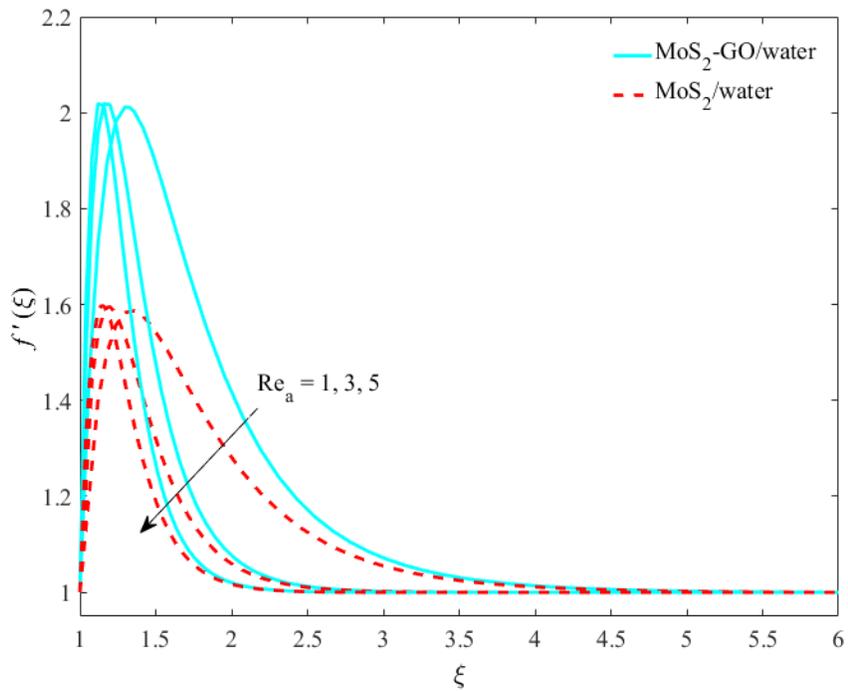

**Fig. 10:** Impact of $m$ on $f'(\xi)$.

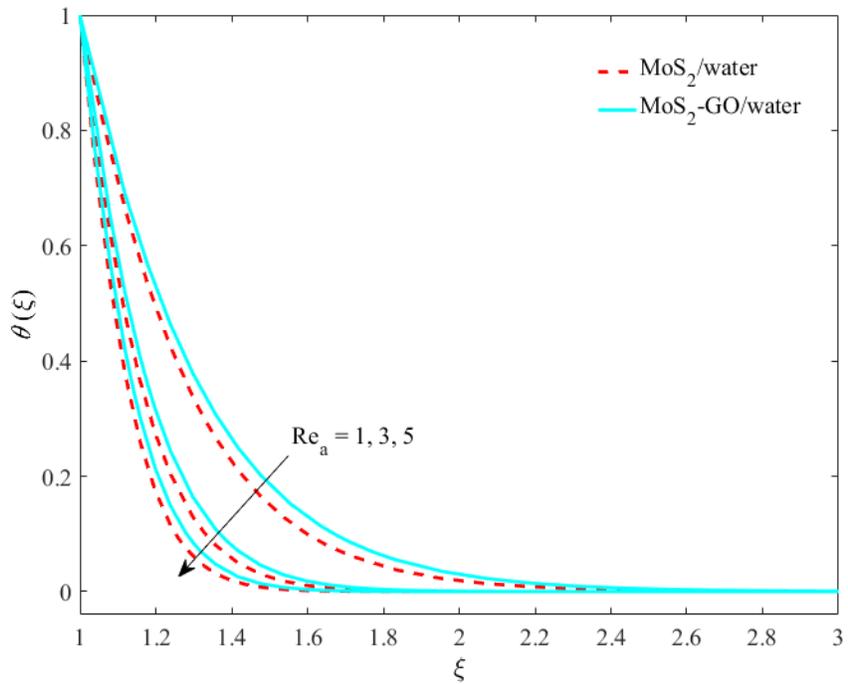

**Fig. 11:** Impact of $m$ on $\theta(\xi)$.

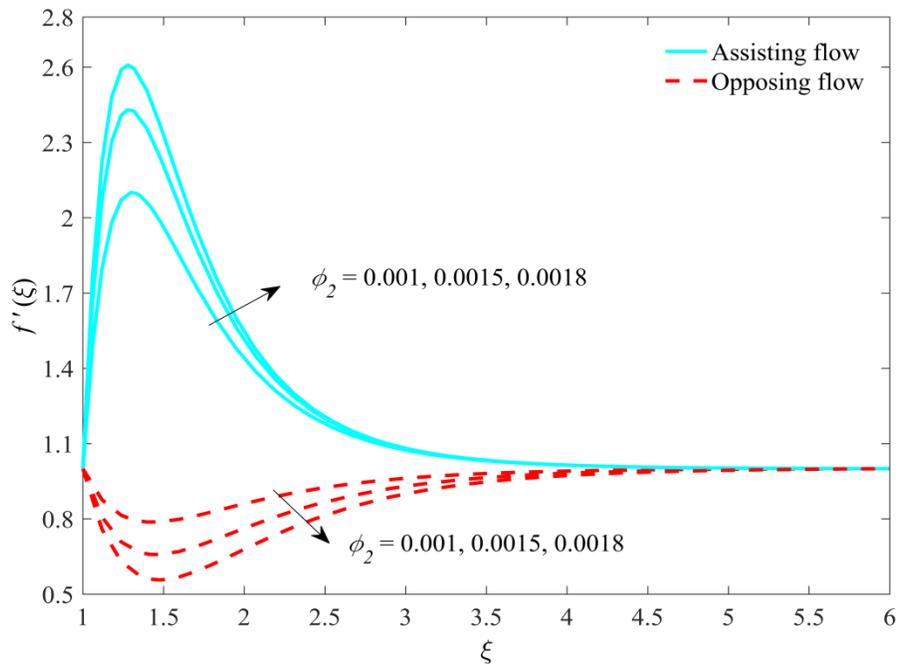

**Fig. 12:** Impact of $\phi_2$ on $f'(\xi)$.

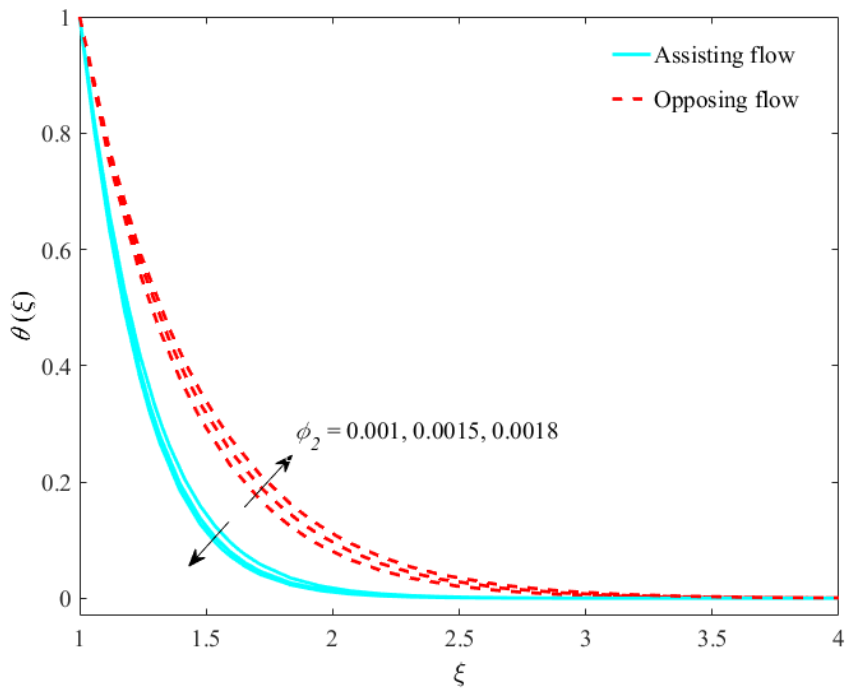

**Fig. 13:** Impact of $\phi_2$ on $\theta(\xi)$.